\documentclass[aps, prb, superscriptaddress, twocolumn]{revtex4}
\usepackage{amssymb}
\usepackage{graphicx}
\usepackage{hyperref}

\begin{document}

\title{Structure and magnetic, thermal, and electronic transport properties of single crystal EuPd$_2$Sb$_2$}
\author{S. Das}
\affiliation{Ames Laboratory and Department of Physics and Astronomy, Iowa State University, Ames, Iowa 50011}
\author{K. McFadden}
\affiliation{Ames Laboratory and Department of Physics and Astronomy, Iowa State University, Ames, Iowa 50011}
\author{Yogesh Singh}
\affiliation{Ames Laboratory and Department of Physics and Astronomy, Iowa State University, Ames, Iowa 50011}
\author{R. Nath}
\affiliation{Ames Laboratory and Department of Physics and Astronomy, Iowa State University, Ames, Iowa 50011}
\author{A. Ellern} 
\affiliation{Department of Chemistry, Iowa State University, Ames, Iowa 50011} 
\author{D. C. Johnston}
\affiliation{Ames Laboratory and Department of Physics and Astronomy, Iowa State University, Ames, Iowa 50011}

\date{\today}
    
\begin{abstract}

Single crystals of EuPd$_2$Sb$_2$ have been grown from PdSb self-flux. The properties of the single crystals have been investigated by x-ray diffraction, magnetic susceptibility $\chi$, magnetization $M$, electrical resistivity $\rho$, Hall coefficient $R_{\rm H}$, and heat capacity $C_{\rm p}$ measurements versus temperature $T$ and magnetic field $H$. Single crystal x-ray diffraction studies confirmed that EuPd$_2$Sb$_2$ crystallizes in the CaBe$_2$Ge$_2$-type structure. The $\chi(T)$ measurements suggest antiferromagnetic ordering at 6.0~K with the easy axis or plane in the crystallographic $ab$ plane. An additional transition occurs at 4.5~K that may be a spin reorientation transition. The $C_{\rm p}(T)$ data also show the two transitions at 6.1~K and 4.4~K, respectively, indicating the bulk nature of the transitions. The 4.4~K transition is suppressed below 1.8~K while the 6.1~K transition moves down to 3.3~K in $H = 8$~T. The $\rho(T)$ data show metallic behavior down to 1.8~K along with an anomaly at 5.5~K in zero field. The anomaly is suppressed to 2.7~K in an 8~T field. The $R_{\rm H}$ measurements indicated that the dominant charge carriers are electrons. The $M(H)$ isotherms show three field-induced transitions at 2.75~T, 3.90~T, and 4.2~T magnetic fields parallel to the $ab$ plane at 1.8~K\@. No transitions are observed in $M(H)$ for fields parallel to the $c$ axis.

\end{abstract}
\maketitle

\section{\label{intro}Introduction} 

\begin{figure}[t]
\includegraphics[width=3.3in]{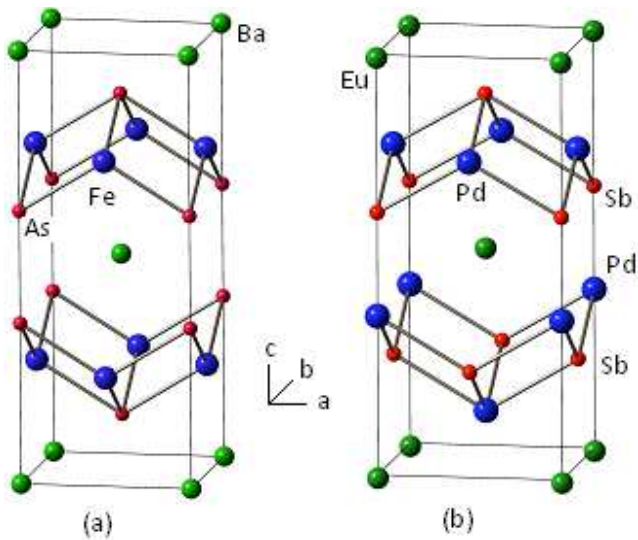}
\caption{(Color online) (a) Crystal structure of BaFe$_2$As$_2$ with the tetragonal ThCr$_2$Si$_2$-type structure. The structure consists of alternating FeAs and Ba layers stacked along the crystallographic $c$ axis. (b) Crystal structure of EuPd$_2$Sb$_2$ with the origin of the unit cell shifted by (1/4 1/4 1/4) compared to that in the space group $P4/nmm$, for comparison purposes. The structure consists of alternating PdSb and Eu layers stacked along the crystallographic $c$ axis similar to the BaFe$_2$As$_2$ shown in (a). However, half of the PdSb layers are inverted (the Pd and Sb atoms switch positions) with respect to the FeAs-type layers.}
\label{struc}
\end{figure}

The recent discovery of high-temperature superconductivity in $R$FeAsO$_{1-x}$F$_x$ ($R$ = La, Ce, Pr, Nd, Sm, Gd, Tb, and Dy)\cite{kamihara,chennature,chen,ren,yang,bos} compounds with superconducting transition temperatures $T_{\rm c}$ as high as 55~K has sparked a lot of interest in the search for new superconductors. These materials crystallize in the tetragonal ZrCuSiAs-type structure with space group $P4/nmm$.\cite{quebe} The structure consists of alternating FeAs and $R$O layers stacked along the crystallographic $c$ axis. The parent compounds $R$FeAsO exhibit spin density wave (SDW) transitions at temperatures $\lesssim 200$~K\@.\cite{chen,dong,klauss} Upon doping with F, the SDW is suppressed and superconductivity appears.\cite{chennature,chen,ren,yang,bos,dong,giovannetti}

Another group of structurally related parent compounds with the chemical formula $A$Fe$_2$As$_2$ ($A$ = Ca, Sr, Ba, and Eu) was soon discovered to show superconductivity upon doping or application of pressure. These compounds crystallize in the tetragonal ThCr$_2$Si$_2$-type structure with space group $I4/mmm$ (No.~139). The structure consists of alternating FeAs and $A$ layers stacked along the $c$ axis as shown in Fig.~\ref{struc}(a). In the FeAs layers, the Fe atoms form a square planar lattice. The $A$Fe$_2$As$_2$ compounds also show SDW and structural transitions at high temperatures\cite{rotter,krellner,ni,yan,ni2,ronning,goldman,tegel,renprb,jeevan} which are suppressed by doping with K, Na, and Cs at the $A$ site and accompanied by the onset of superconductivity.\cite{rotterprl,chen2,jeevan2,sasmal}

In both classes of $R$FeAsO$_{1-x}$F$_x$ and $A$Fe$_2$As$_2$ compounds described above, FeAs layers that are stacked along the $c$ axis are evidently a key building block yielding superconductors with relatively high $T_{\rm c}$. This gives a strong motivation to investigate similarly structured compounds in a search for additional high-$T_{\rm c}$ superconductors. 

The compound EuPd$_2$Sb$_2$ crystallizes in the CaBe$_2$Ge$_2$-type structure with space group $P4/nmm$ (No.~129),\cite{hofmann} as shown in Fig.~\ref{struc}(b). The structure is closely related to the $A$Fe$_2$As$_2$ structure. Alternating PdSb and Eu layers are stacked along the $c$ axis, similar to the $A$Fe$_2$As$_2$ structure. However, there is a distinct difference between the two structures. In half of the PdSb layers in the EuPd$_2$Sb$_2$ structure, the Pd atoms are arranged in a planar square lattice with two Sb layers on either side of each Pd layer, resulting in a tetrahedral coordination of Pd by Sb as in the FeAs-type layers. However, alternating with these layers are layers in which the Pd and Sb positions are switched, as shown in Fig.~\ref{struc}(b). 

There have been reports of structural instabilities and antiferromagnetic ordering in some compounds forming in the CaBe$_2$Ge$_2$-type structure. UCu$_{1.5}$Sn$_2$ orders antiferromagnetically at 110 $^\circ$C which is very high among uranium intermetallics.\cite{purwanto1996} CePd$_2$Ga$_2$ undergoes a tetragonal to monoclinic second order structural transition at 125~K and orders antiferromagnetically at 2.3~K\@.\cite{kitagawa1999} LaPd$_2$Ga$_2$ is superconducting below 1.9~K\@. \cite{kitagawa1999} Eu was reported to be in a mixed valent state between Eu$^{+2}$ (spin $S = 7/2$) and Eu$^{+3}$ (spin $S = 0$) in polycrystalline samples of EuPd$_2$Sb$_2$.\cite{hofmann} In this paper, we report the synthesis and structure of single crystals of EuPd$_2$Sb$_2$ and their physical properties including magnetic susceptibility, magnetization, specific heat, and electronic transport measurements.

\section{\label{expt}Experimental Details}

\begin{figure}[t]
\includegraphics[width=2.5in]{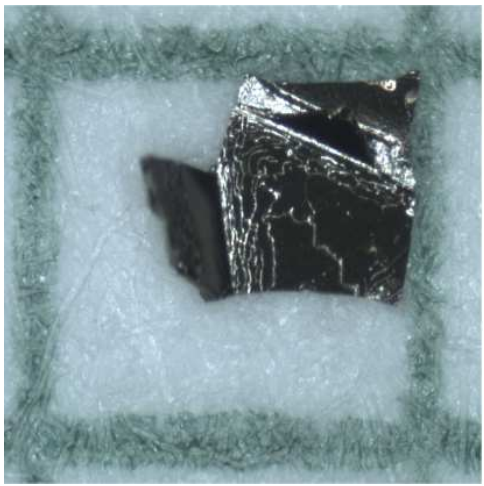}
\caption{An as-grown crystal of EuPd$_2$Sb$_2$. The grid size in 1 mm.}
\label{pic}
\end{figure}

Single crystals of EuPd$_2$Sb$_2$ were grown using PdSb self-flux which melts at $\sim$~805~$^\circ$C\@. The Eu (99.999\% pure) was obtained from the Ames Laboratory Materials Preparation Center.  The Pd (99.95\% pure) and Sb (99.999\% pure) were obtained from Alfa-Aesar. Pd and Sb powders were thoroughly mixed inside a helium-filled glove box, and then poured on top of a chunk of Eu ($\sim$~0.1 g) that was placed at the bottom of a 2~mL alumina crucible.  The elements were in the atomic ratio Eu:Pd:Sb = 1:5:5. The top of the crucible was packed with quartz wool. The crucible was then sealed in a quartz tube under vacuum and was placed vertically in a box furnace and heated to 1000 $^\circ$C at a rate of 76 $^\circ$C/h and held there for 6 h. The tube was then cooled to 850 $^\circ$C at the rate of 1.5 $^\circ$C/h and at this temperature the tube was removed from the oven and centrifuged to partially separate the flux from the crystals. A single conglomerated chunk (about 0.4~g) was found in the crucible after removing the quartz wool. Plate-like gold-colored crystals were isolated mechanically. The largest crystals had dimensions $\sim$~2$\times$2$\times$0.1~mm$^3$. The crystals are brittle and are easily broken into smaller pieces. Figure~\ref{pic} shows an as-grown crystal on a mm grid.

Single crystal x-ray diffraction measurements were done using a Bruker CCD-1000 diffractometer with Mo K$_{\alpha}$ ($\lambda$ = 0.71073 \AA) radiation. Magnetic measurements on the crystals were carried out using a Quantum Design superconducting quantum interference device (SQUID) magnetometer in the temperature $T$ range 1.8--350~K and magnetic field $H$ range 0--5.5~T\@. Heat capacity, electrical resistivity, and Hall coefficient measurements were done using a Quantum Design physical property measurement system (PPMS). For the heat capacity measurements, Apiezon N grease was used for thermal coupling between a sample and the sample platform. The heat capacity was measured in the temperature range 1.8--300~K in $H$ = 0, 2, 5, 7, and 9~T\@. For electrical resistivity and Hall coefficient measurements, platinum leads were attached to the crystals using silver epoxy. Electrical resistivity measurements were carried out using the standard AC four probe method with 10~mA excitation current in the temperature range 1.8--300~K and magnetic field range 0--8~T\@. Hall coefficient measurements were carried out using the five-wire configuration supported by the PPMS ACT\cite{ppms} option with 100~mA excitation current in the temperature range 1.8--310~K and magnetic field range 0--8~T\@. The Hall voltage was computed at each temperature from the odd part of the measured transverse voltage upon reversing the sign of the applied magnetic field. The even part was much smaller than the odd part at each measured temperature.

\section{\label{results}Results}

\subsection{\label{xray}Structure and chemical composition determination}

\begin{table}
\caption{Crystal data and structure refinement of ${\rm EuPd_2Sb_2}$ at a temperature of 173~K\@. Here R1 = $\sum$$\mid$$\mid$$F$$_{\rm obs}$$\mid$~$-$~$\mid$$F$$_{\rm calc}$$\mid$$\mid$/$\sum$$\mid$$F$$_{\rm obs}$$\mid$ and wR2 = ($\sum$[ $w$($\mid$$F$$_{\rm obs}$$\mid$$^2$ $-$ $\mid$$F$$_{\rm calc}$$\mid$$^2$)$^2$]/$\sum$[ $w$($\mid$$F$$_{\rm obs}$$\mid$$^2$)$^2$])$^{1/2}$, where $F$$_{\rm obs}$ is the observed structure factor and $F$$_{\rm calc}$ is the calculated structure factor.}

\begin{ruledtabular}
\begin{tabular}{ll}
Crystal system/Space group & Tetragonal, $P4/nmm$\\
Unit cell parameters & $a$ = 4.653(2) \AA\\
                     & $c$ = 10.627(4) \AA\\
Unit cell volume & 230.1(3) \AA$^3$\\
$Z$ (formula units/unit cell) & 2\\
Density (Calculated) & 8.779 Mg/m$^3$\\
Absorption coefficient & 32.47 mm$^{-1}$\\
$F$(000) & 514\\
Goodness-of-fit on $F$$^2$ & 1.235\\
Final $R$ indices [$I$ $>$ 2$\sigma$($I$)] & R1 = 0.0737\\
& wR2 = 0.02506\\ 
Extinction coefficient & 0.033(9) mm$^{-1}$\\
\end{tabular}
\end{ruledtabular}
\label{parameters}
\end{table}

\begin{table}
\caption{Atomic coordinates $x$, $y$, and $z$ (10$^{-4}$) and equivalent isotropic displacement parameters $U$ (10$^{-3}$~\AA$^2$) for ${\rm EuPd_2Sb_2}$ at 173 K\@.}
\begin{ruledtabular}
\begin{tabular}{lllll}
 & $x$ & $y$ & $z$ & $U$(eq)\\
\hline
Eu & 2500 & 2500 & 2425(1) & 13(1)\\
Pd(1) & 7500 & 2500 & 0 & 16(1)\\
Pd(2) & 2500 & 2500 & 6292(2) & 17(1)\\
Sb(1) & 7500 & 2500 & 5000 & 13(1)\\
Sb(2) & 2500 & 2500 & 8738(1) & 14(1)\\
\end{tabular}
\end{ruledtabular}
\label{thermal}
\end{table} 

A well-shaped crystal with dimensions $0.21\times0.18\times0.11$ ${\rm mm}^3$ was selected for single crystal x-ray diffraction at 173~K\@. X-ray structure determination and refinement were performed using the SHELXTL software package.\cite{shelxtl} The refined unit cell parameters, the isotropic thermal parameters, and the atomic positions are listed in Tables~\ref{parameters} and \ref{thermal}. Our results confirm that EuPd$_2$Sb$_2$ crystallizes in the CaBe$_2$Ge$_2$ structure.\cite{hofmann} The unit cell dimensions and the atomic positions are similar to those found from single crystal x-ray diffraction measurements at room temperature in Ref.~\onlinecite{hofmann}, which were $a$ = 4.629(1)~\AA, $c$ = 10.568(2)~\AA, Eu: $z$ = 0.2424(1); Pd(2): $z$ = 0.6284(2); Sb(2): $z$ = 0.8745(1). The significant difference between the lattice parameters in Ref.~\onlinecite{hofmann} and lattice parameters obtained by us suggests a difference in crystal stoichiometry between the samples in Ref.~\onlinecite{hofmann} and ours, although both studies indicate nearly stoichiometric compositions. The temperature difference between the two studies cannot be responsible, since the lattice parameter differences are opposite to expectation in that case.

The stoichiometry of a representative crystal was checked by semiquantitative energy-dispersive x-ray (EDX) microanalysis. The results gave the following composition: Eu, 24.9~$\pm$~1.1~wt$\%$; Pd, 35.5~$\pm$~0.8~wt$\%$; Sb, 39.7~$\pm$~1.0~wt$\%$. These values are consistent with the values calculated for the composition EuPd$_2$Sb$_2$: Eu, 24.98~wt$\%$; Pd, 34.98~wt$\%$; Sb, 40.03~wt$\%$.

\subsection{\label{magnetic}Magnetic measurements}

\subsubsection{\label{susceptibility}Magnetic susceptibility measurements}

\begin{figure}[t]
\includegraphics[width=2.5in]{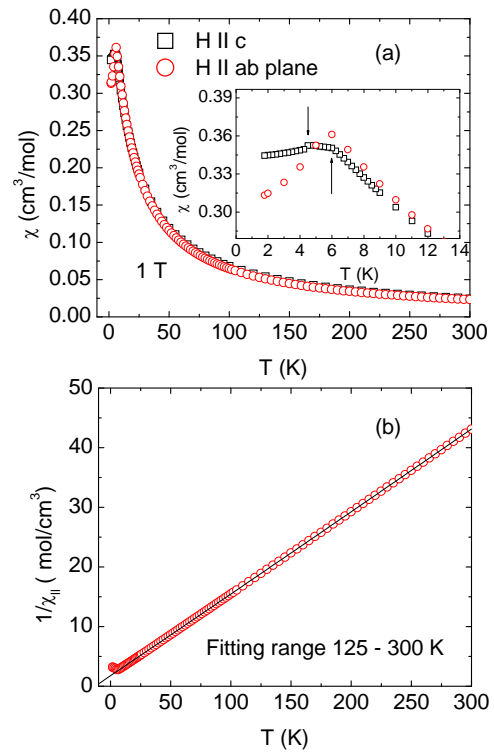}
\caption{(Color online) (a) Magnetic susceptibility $\chi$ versus temperature $T$ of EuPd$_2$Sb$_2$ with the magnetic field $H$ parallel to the crystallographic $c$ axis and to the $ab$ plane, respectively. The inset shows the low-$T$ part of the $\chi(T)$ plot. The two transitions at 4.5~K and 6.0~K are indicated by the vertical arrows. (b) Inverse susceptibility 1/$\chi(T)$ for $H \parallel c$. The solid curve is the Curie-Weiss fit [Eq.~(\ref{cw})] to the 1/$\chi_{\parallel}(T)$ data in the temperature range 125--300~K with the parameters listed in the text.}
\label{susc}
\end{figure}

Figure~\ref{susc}(a) shows the magnetic susceptibility $\chi$ of EuPd$_2$Sb$_2$ versus temperature $T$ with the magnetic field $H$ parallel to the crystallographic $c$ axis ($\chi_{\parallel}$) and to the $ab$ plane ($\chi_{\perp}$), respectively. At high-$T$, the $\chi(T)$ shows nearly isotropic paramagnetic behavior. Figure~\ref{susc}(b) shows the inverse susceptibility $1/\chi$ for $H \parallel c$ versus $T$. An excellent fit to the data in the $T$ range 125 -- 300~K was obtained using the Curie-Weiss behavior  
\begin{equation}
\frac{1}{\chi} = \frac{1}{\chi_0 + C/(T - \theta)},
\label{cw}
\end{equation}
where $\chi_0$ is the $T$-independent susceptibility, $C$ is the Curie constant, and $\theta$ is the Weiss temperature. The values of the parameters obtained from the fit are $C = 7.333(8)$~cm$^3$~K/mol, $\theta = -12.9(2)$~K, and $\chi_0 = -0.00024(3)$~cm$^3$/mol. Keeping $\chi_0$ fixed to zero, the Curie-Weiss fits to the 1/$\chi_{\parallel}(T)$ data in the different temperature ranges between 25--300~K and 200--300~K yielded $C$~=~7.23(3)~cm$^3$~K/mol and $\theta =-11.8(8)$~K\@. The obtained Curie constants are significantly lower than the value $C = 7.88$ cm$^3$ K/mol expected for Eu$^{+2}$ (spin $S = 7/2$) with $g$-factor $g = 2$. This indicates that Eu is in an intermediate valent state Eu$^{+2.07}$ as previously suggested in Ref.~\onlinecite{hofmann}. The negative Weiss temperature indicates  dominant antiferromagnetic interactions between the nearest-neighbor Eu spins. 

At low temperatures, $\chi_{\parallel}$ becomes almost $T$-independent below 6.0~K with a cusp at $T = 4.5$~K as shown in the inset of Fig.~\ref{susc}(a). $\chi_{\perp}$ shows a peak at 6.0~K and decreases monotonically at lower $T$. The data suggest antiferromagnetic ordering of the Eu spins at 6.0~K with the easy axis or plane within the $ab$ plane, with a possible spin reorientation transition at 4.5~K\@.

\subsubsection{\label{magnetization}Magnetization versus applied magnetic field isotherm measurements}

\begin{figure}[t]
\includegraphics[width=3.3in]{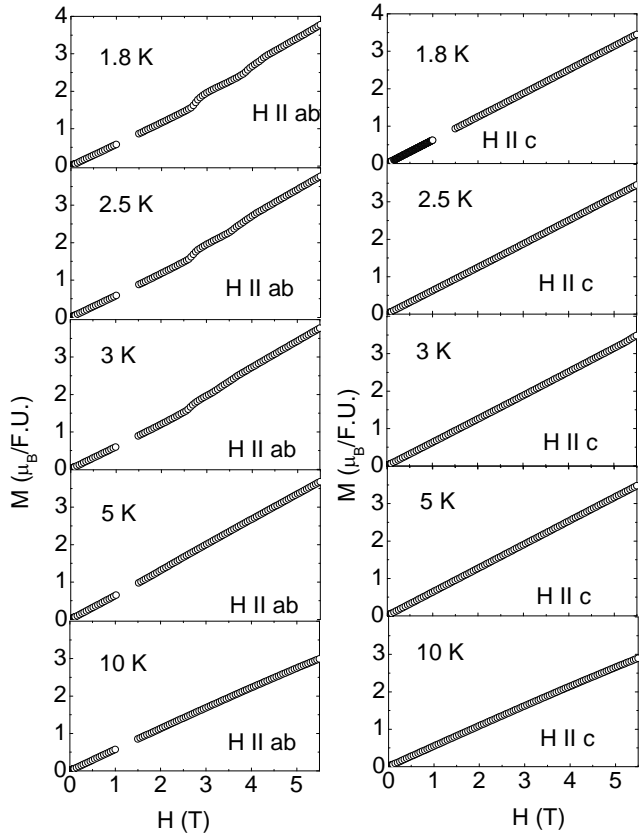}
\caption{Magnetization $M$ versus applied magnetic field $H$ of EuPd$_2$Sb$_2$ with $H$ parallel to the crystallographic $c$ axis (left-hand panels) and to the $ab$ plane (right-hand panels), respectively.}
\label{mh}
\end{figure}

\begin{figure}[t]
\includegraphics[width=2.5in]{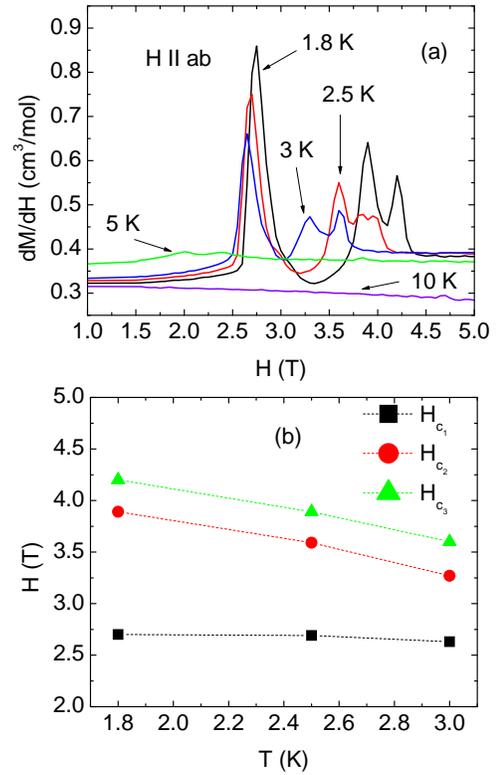}
\caption{(a) Derivative $dM/dH$ of the magnetization $M$ with respect to the applied field $H$ versus $H$ with $H$ parallel to the crystallographic $ab$ plane. (b) The fields $H_{c_1}$, $H_{c_2}$, and $H_{c_3}$, at which transitions are observed in $dM/dH$, versus $T$. The dotted lines are guides to the eye.}
\label{mhderiv}
\end{figure}

Figure~\ref{mh} shows the magnetization $M$ of EuPd$_2$Sb$_2$ versus magnetic field $H$ with $H$ parallel to the crystallographic $c$ axis (right-hand panels) and to the $ab$ plane (left-hand panels), respectively. For $H \parallel ab$, anomalies in $M(H)$ are clearly visible for $T < 5$~K\@. Above 10~K, $M(H)$ is proportional to $H$. To illustrate the anomalies more clearly, Fig.~\ref{mhderiv}(a) shows the derivative $dM/dH$ versus $H$ with $H \parallel ab$. The $dM/dH$ data for $M \parallel ab$ show three peaks at $H_{c_1} = 2.75$~T, $H_{c_2} = 3.90$~T, and $H_{c_3} = 4.20$~T, respectively, at 1.8~K\@. The temperature dependences of the fields at which these field-induced transitions occur are shown in Fig.~\ref{mhderiv}(b). The transition fields are seen to decrease with increasing $T$, and disappear between 5 and 10~K\@. At 1.8~K in $H$ = 5.5~T, the value of $M \parallel ab$ in Fig.~\ref{mh} is 3.8 $\mu_{\rm B}$/f.u. This value is much less than the expected Eu$^{+2}$ saturation moment of 7 $\mu_{\rm B}$. This difference suggests that the metamagnetic transitions take place between different antiferromagnetic states. In contrast, $M \parallel c$ is proportional to $H$ at all $T$\@. Qualitatively similar $M(H)$ observations were previously reported for single crystals of EuRh$_2$As$_2$.\cite{singh2009}

\subsection{\label{heatcapacity}Heat capacity measurements}

\begin{figure}[t]
\includegraphics[width=2.5in]{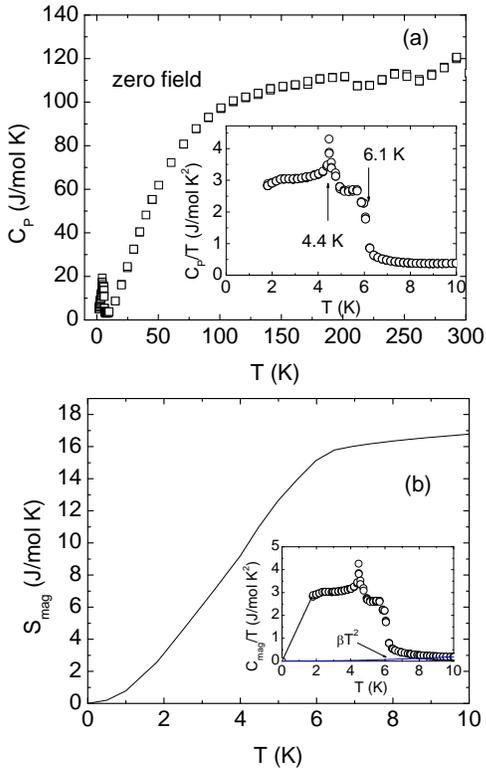}
\caption{(Color online) (a) Heat capacity $C_{\rm p}$ versus temperature $T$ of EuPd$_2$Sb$_2$ in zero magnetic field. The inset shows $C_{\rm p}/T$ versus $T$ for $T < 10$~K\@. Two anomalies in $C_{\rm p}(T)/T$ are observed at 4.4~K and 6.1~K indicated by vertical arrows in the inset. (b) Calculated magnetic entropy $S_{\rm mag} = \int\limits_{0}^{T}[C_{\rm mag}(T)/T] dT$ versus $T$. A linear extrapolation to zero of $C_{\rm p}(T)/T$ as shown by the dotted straight line in the inset was used to approximate the missing $C_{\rm p}/T$ data between 0~K and 1.8 K\@. The solid line in the inset is a plot of the lattice contribution $C_{\rm latt} = \beta T^2$ with $\beta = 1.92$~mJ/mol~K$^4$ obtained for BaRh$_2$As$_2$ in Ref.~\onlinecite{singh2008}. The inset shows that the lattice heat capacity is negligible compared to the magnetic heat capacity below 10~K\@.}
\label{hc}
\end{figure}

\begin{figure}[t]
\includegraphics[width=2in]{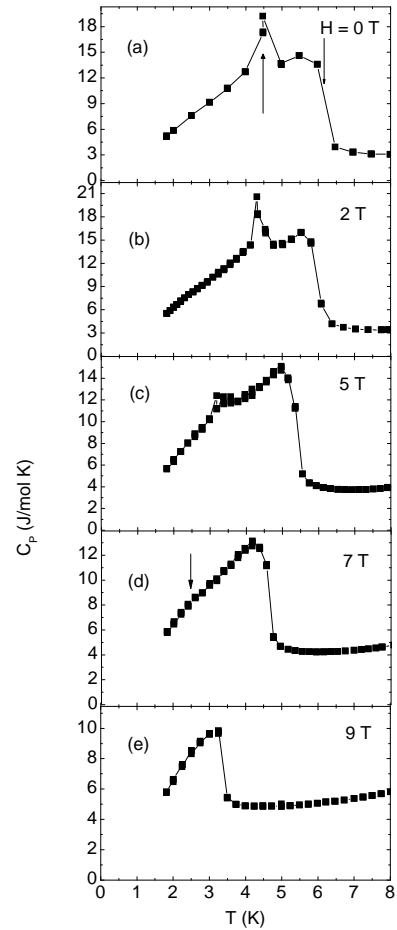}
\caption{Heat capacity $C_{\rm p}$ versus temperature $T$ of EuPd$_2$Sb$_2$ in different magnetic fields parallel to the $c$ axis. The two transitions at temperatures ${T_{\rm N}}_1$ and ${T_{\rm N}}_2$, respectively, are indicated in panel (a). The vertical arrow in (d) points to $T_{\rm N_2}$ in $H = 7$~T\@.}
\label{hcT}
\end{figure}

\begin{figure}[t]
\includegraphics[width=2.5in]{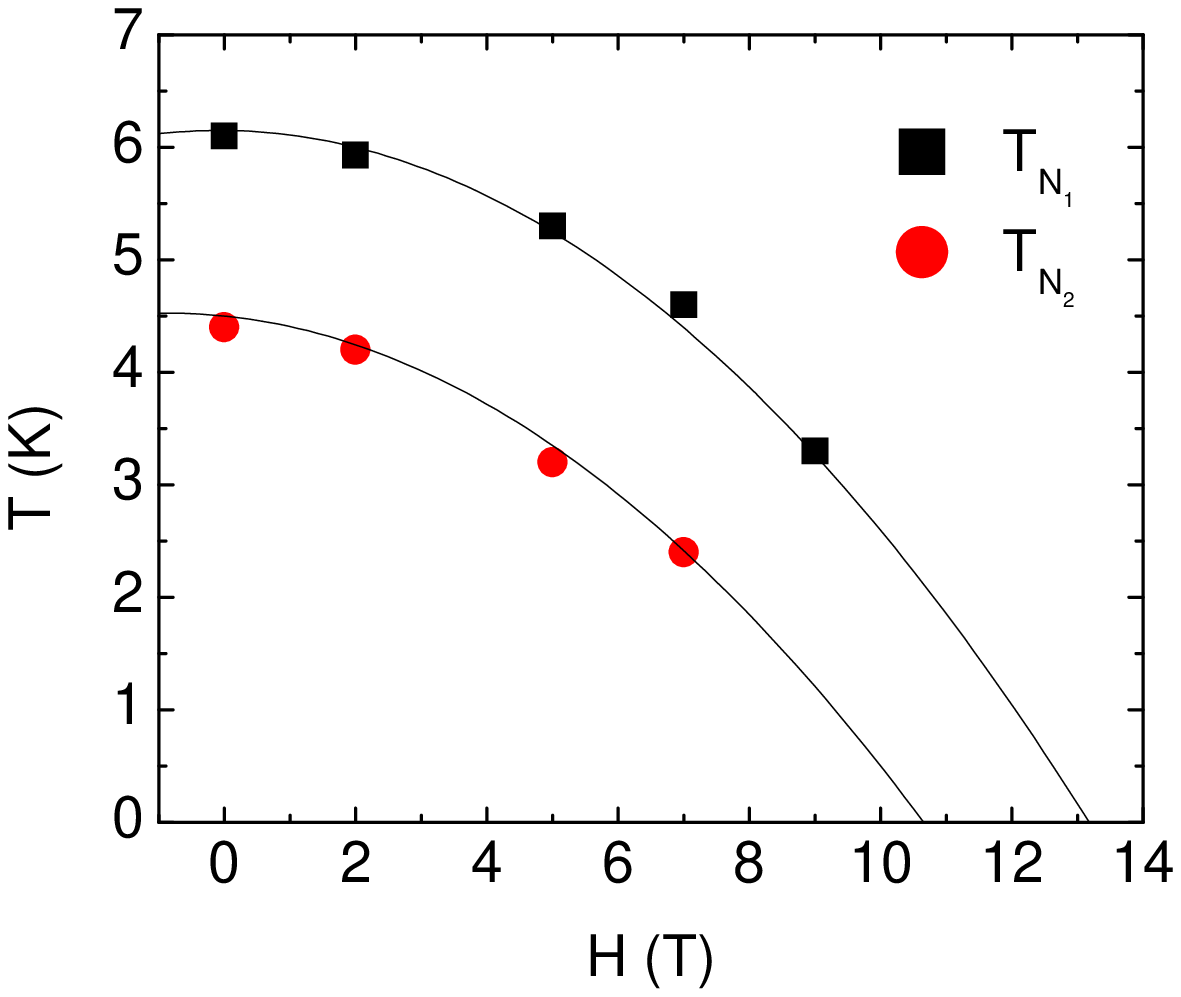}
\caption{(Color online) Transition temperatures $T_{\rm N_1}$ and $T_{\rm N_2}$ of EuPd$_2$Sb$_2$ versus magnetic field $H \parallel c$ as determined from the heat capacity measurements versus $H$ in Fig.~\ref{hcT}. The solid curves are guides to the eye.}
\label{tcT}
\end{figure}

Figure~\ref{hc}(a) shows the heat capacity $C_{\rm p}$ of a 2.619~mg EuPd$_2$Sb$_2$ crystal versus temperature $T$ in zero magnetic field. The inset of Fig.~\ref{hc}(a) shows $C_{\rm p}/T$ versus $T$ for $T < 10$~K\@. Two anomalies are observed at 6.1~K and 4.4~K, respectively, indicating that the transitions observed in $\chi(T)$ in the inset of Fig.~\ref{susc}(a) at similar temperatures are bulk long-range magnetic ordering transitions. The data at high $T$ $\sim 300$~K approach the Dulong-Petit classical lattice heat capacity value of 15$R \simeq 125$~J/mol~K, where $R$ is the molar gas constant. 

Figure~\ref{hc}(b) shows the calculated magnetic entropy $S_{\rm mag} = \int\limits_{0}^{T}[C_{\rm mag}(T)/T] dT$ versus $T$ at low temperatures $T < 10$~K, where $C_{\rm mag} = C_{\rm p} - C_{\rm latt}$ is the magnetic contribution and $C_{\rm latt}$ is the lattice contribution to the specific heat. We assumed $C_{\rm latt}$~=~$\beta T^3$ for $T < 10$~K with $\beta~=~1.93(4)$~mJ/mol~K$^4$ obtained for BaRh$_2$As$_2$ from Ref.~\onlinecite{singh2008}. A linear extrapolation to zero of $C_{\rm mag}/T$, as shown by the dotted straight line in the inset of Fig.~\ref{hc}(b), was assumed in order to approximate the missing $C_{\rm mag}/T$ data between 0~K and 1.8~K\@. The magnetic entropy $S_{\rm mag} = 16.4$~J/mol~K at 10~K is close to the expected entropy $S_{\rm mag} = R{\rm ln}(2S+1)$~=~17.3~J/mol~K due to ordering of one spin $S = 7/2$ per formula unit.

 Figures~\ref{hcT}(a)--(e) show $C_{\rm p}(T)$ in different magnetic fields parallel to the crystallographic $c$ axis. For $H = 0$~T, $C_{\rm p}(T)$ shows a jump at ${T_{\rm N}}_1 = 6.1$~K and then a cusp at ${T_{\rm N}}_1 = 4.4$~K\@. The shapes of the $C_{\rm p}$ anomalies at the two transitions are thus distinctly different. As $H$ is increased, $T_{\rm N_2}$ decreases below 1.8~K at 9~T, while the $T_{\rm N_1}$ goes down to 3.2~K in 9~T\@. The transition at $T_{\rm N_1}$ remains sharp while the transition at $T_{\rm N_2}$ broadens for $H > 2$T\@. Figure~\ref{tcT} shows plots of $T_{\rm N_1}$ and $T_{\rm N_2}$ versus $H$.

\subsection{\label{trans}Electronic transport measurements}

\subsubsection{\label{resistivity}Electrical resistivity measurements}

\begin{figure}[t]
\includegraphics[width=2.5in]{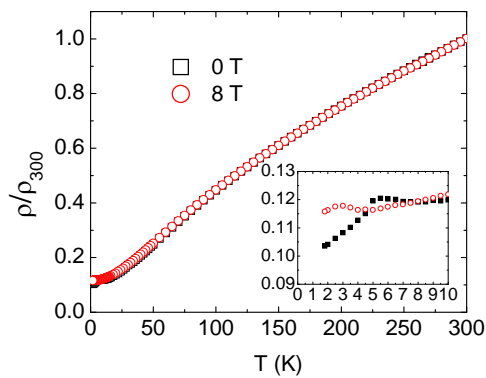}
\caption{(Color online) Electrical resistance ratio $\rho$/$\rho_{300}$ versus temperature $T$ of EuPd$_2$Sb$_2$ for current in the $ab$-plane in 0 and 8~T magnetic fields parallel to the $c$ axis where $\rho$ is the resistance at temperature $T$ and $\rho_{300} = (50\pm25$) $\mu\Omega$ cm is the resistance at 300~K\@. The inset shows the low-$T$ region below 10~K\@.}
\label{res}
\end{figure}

Figure~\ref{res} shows the electrical resistance ratio $\rho$/$\rho_{300}$ of EuPd$_2$Sb$_2$ for current parallel to the $ab$-plane versus temperature $T$ in 0 and 8~T magnetic fields parallel to the $c$ axis, where $\rho$ is the resistance at temperature $T$ and $\rho_{300} = (50\pm25) \mu\Omega$ cm is the resistance at 300~K\@. The large fractional uncertainty in $\rho_{300}$ arises due to the uncertainty in the geometric factor for the irregularly-shaped crystal. The inset shows the low-$T$ region below 10~K\@. The resistance data exhibit metallic behavior down to the lowest temperature. The residual resistance ratio $RRR = \rho(300~\rm K)/\rho(2~\rm K) \approx 10$. This value is comparable to the values found in the $ab$-plane resistivity for single crystals of other layered pnictides.\cite{rotter,krellner,renprb,ni2} From the inset of Fig.~\ref{res}, in zero magnetic field $\rho(T)$ shows an anomaly at 5.4~K which gets suppressed to 2.9~K in $H = 8$~T\@. The anomaly is evidently due to the antiferromagnetic ordering at $T_{N_1} = 6.1$~K as observed from the $C_{\rm p}(T)$ and $\chi(T)$ measurements.

\subsubsection{\label{halleffect}Hall coefficient measurements}

\begin{figure*}[t]
\includegraphics[width=4in]{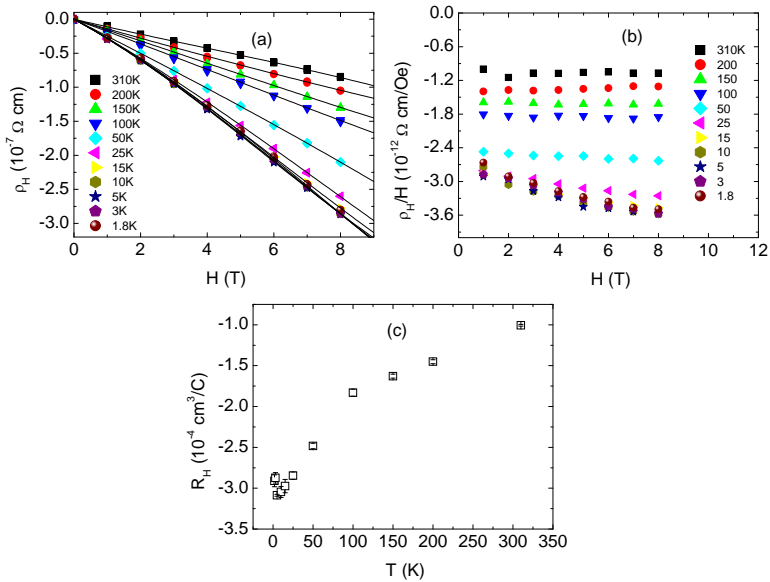}
\caption{(Color online) (a) Hall resistivity $\rho_{\rm H}$ of EuPd$_2$Sb$_2$ versus applied magnetic field $H$ at the indicated temperatures $T$. The solid curves are polynomial fits to the data (see text). (b) $\rho_{\rm H}/H$ versus $H$ at the indicated values of $T$. (c) Hall coefficient $R_{\rm H}$ versus $T$. The consistently negative $R_{\rm H}$ indicates that the dominant current carriers are electrons.}
\label{hall}
\end{figure*}

Figure~\ref{hall}(a) shows the Hall resistivity $\rho_{\rm H} = V_{\rm H} A/Il$ versus $H$ where $V_{\rm H}$ is the measured Hall voltage, $A$ is the cross sectional area of the sample, $l$ is the separation of the transverse voltage leads, and $I$ is the longitudinal current. In Fig.~\ref{hall}(a), $\rho_{\rm H}$ versus magnetic field $H$ is seen to deviate from a proportional behavior below 100~K\@. This behavior is more clearly seen in the plot of $\rho_{\rm H}$/$H$ versus $H$ in Fig.~\ref{hall}(b). The measured $\rho_{\rm H}(H)$ data were fitted by the function $\rho_{\rm H}(H) = a_1 H + a_3 H^3 + a_5 H^5$, as shown by the solid curves in Fig.~\ref{hall}(a), and $a_1$ (the coefficient of the linear term) is the Hall coefficient $R_{\rm H}$. $R_{\rm H}$ versus $T$ is shown in Fig.~\ref{hall}(c), where $R_{\rm H}$ becomes more negative by a factor of 3 on cooling from 310~K to 2~K\@. The temperature dependence is very similar to $R_{\rm H}(T)$ of BaRh$_2$As$_2$ (Ref.~\onlinecite{singh2008}) which crystallizes in the tetragonal ThCr$_2$Si$_2$-type structure. The consistently negative $R_{\rm H}$ indicates that the dominant charge carriers are electrons. If one uses a single band model, one obtains a conduction electron concentration $n = (R_{\rm H}e)^{-1} = 11$ and 3 (f.u.)$^{-1}$ at 310~K and 2~K, respectively.

\section{\label{summary}Summary and discussion}

We have synthesized single crystals of EuPd$_2$Sb$_2$ and characterized them using single crystal x-ray diffraction, anisotropic magnetic susceptibility and magnetization, specific heat, electrical resistivity, and Hall coefficient measurements. The magnetic susceptibility indicates antiferromagnetic ordering at 6.0~K with an easy axis or plane within the crystallographic $ab$ plane followed by another transition at 4.5~K\@. The transitions are also observed in heat capacity measurements indicating their bulk nature. The transition at 4.5~K is suppressed below 1.8~K in a magnetic field of 8~T as observed from the heat capacity and electrical resistivity measurements. The transition at 6~K is pushed down to 3.5~K in a field of 8~T\@. $M(H)$ isotherms show three field-induced transitions at 2.75~T, 3.90~T, and 4.2~T for magnetic fields parallel to the $ab$ plane at 1.8~K\@. No transitions are observed for fields parallel to the $c$ axis. The Hall coefficient is consistently negative from 1.8 to 310~K indicating electrons as the dominant charge carriers.

\begin{table*}
\caption{A comparison of the structural and magnetic parameters of EuPd$_2$Sb$_2$ with some Eu compounds which form in the related ThCr$_2$Si$_2$-type structure. The structure type and the lattice parameters $a$ and $c$ are at room temperature for EuFe$_2$As$_2$ and EuRh$_2$As$_2$, at 124~K for EuNi$_2$As$_2$, and at 173~K for EuPd$_2$Sb$_2$. $T_{\rm N}$ is the antiferromagnetic ordering temperature and $\mu_{\rm eff}$ is the effective moment in the Curie-Weiss law calculated from magnetic susceptibility measurements.}
\begin{ruledtabular}
\begin{tabular}{lllllll}
Compound & Structure-type & $a$~(\AA) \ & $c$~(\AA) \ & $T_{\rm N}$~(K) & $\mu_{\rm eff}$ ($\mu_{\rm B}$/f.u.) & Ref.\\
\hline
EuFe$_2$As$_2$ & ThCr$_2$Si$_2$ & 3.9104 & 12.1362 & 20 & 7.79 & \onlinecite{renprb}\\
EuRh$_2$As$_2$ & ThCr$_2$Si$_2$ & 4.075 & 11.295 & 47 &  & \onlinecite{singh2009}\\
EuNi$_2$As$_2$ & ThCr$_2$Si$_2$ & 4.0964 & 10.029 & 14 &  & \onlinecite{bauer2008}\\
EuPd$_2$Sb$_2$ & CaBe$_2$Ge$_2$ & 4.653 & 10.627 & 6.1 & 7.65 & This work\\
\end{tabular}
\end{ruledtabular}
\label{comparison}
\end{table*} 

A comparison of structural and magnetic parameters of EuPd$_2$Sb$_2$ with those of some Eu compounds which form in the related ThCr$_2$Si$_2$-type structure is given in Table~\ref{comparison}. Of the compounds listed in Table~\ref{comparison}, only EuFe$_2$As$_2$ shows superconducting behavior under pressure\cite{terashima2009} as well as under doping at the Eu site.\cite{jeevan2} $T_{\rm N}$ for EuPd$_2$Sb$_2$ with $a = 4.653$~\AA \ and a $c/a$ ratio of 2.28 is 6.1~K compared to 20~K for EuFe$_2$As$_2$ which has a smaller $a = 3.910$~\AA \ and a much larger $c/a$ = 3.10. Thus the significant difference in $T_{\rm N}$ values is probably due at least in part to these structural differences. In Eu$_{0.5}$K$_{0.5}$Fe$_2$As$_2$ which is superconducting below 32~K, however, AF ordering of the Eu spins still takes place below 10~K\@. The calculated effective moment of the Eu spins in EuPd$_2$Sb$_2$ is close to that of Eu spins in EuFe$_2$As$_2$. The Hall coefficient of  EuPd$_2$Sb$_2$ remains negative between 1.8 -- 300~K like that in the superconducting Ba(Fe$_{1-x}$Co$_x$)$_2$As$_2$ and Ba(Fe$_{1-x}$Cu$_x$)$_2$As$_2$.\cite{mun2009} However, in EuFe$_2$As$_2$, the Hall coefficient changes sign from negative to positive at $\sim175$~K\@. At 300~K, $R_{\rm H} \sim 0$ in EuFe$_2$As$_2$,\cite{renprb} which suggests that the charge carriers comprise both electrons and holes. Probably, the magnetic nature of the FeAs layers in EuFe$_2$As$_2$ is an important factor behind the superconducting behavior at high pressures and low temperatures. It will be very interesting to grow single crystals of EuPd$_{2-x}$Fe$_x$Sb$_2$ and study their physical properties. The Fe-doping at the Pd site will eventually make the Pd(Fe)-Sb layers structurally similar to the FeAs layers in EuFe$_2$As$_2$.

\begin{acknowledgments}
Work at the Ames Laboratory was supported by the Department of Energy-Basic Energy Sciences under Contract No. DE-AC02-07CH11358.
\end{acknowledgments}


\begin{thebibliography} {99}
\bibitem{shelxtl} All software and sources of the scattering factors are contained in the SHELXTL (version 5.1) program library (G. Sheldrick, Bruker Analytical X-Ray Systems, Madison, WI).
\end{thebibliography}
\end{document}